\documentclass[preprint,fleqn,showpacs,showkeys]{revtex4}
\usepackage{graphicx}
\usepackage{amssymb}
\usepackage{amsmath}
\usepackage{bm}
\usepackage{feynmf}
\usepackage{tikz}
\usetikzlibrary{trees}
\usetikzlibrary{decorations.pathmorphing}
\usetikzlibrary{decorations.markings}
\usetikzlibrary{snakes}

\def\be{\begin{equation}}
\def\ee{\end{equation}}
\def\bea{\begin{eqnarray}}
\def\eea{\end{eqnarray}}

\begin{document}
\setcounter{page}{1}
\title{Scalar and vector one-loop massive tadpole light-front gauge Feynman integrals}
\author{Alfredo Takashi \surname{Suzuki}}
\affiliation{Department of Physics, La Sierra University, Riverside, CA 92505}
\author{Timothy \surname{Suzuki}}
\affiliation{Department of Physics, Southern Adventist University, Collegedale, TN 37315}
\date[]{}

\begin{abstract}

Light-front gauge is the most popular one to work with fundamental interactions, due to its characteristic maximum kinematical Poincare operators that it allows. However, it is also known to be one of the trickiest gauges one can work with for gauge theories, due to its singular nature. So, in terms of perturbative calculations in the light-front, there are only a few published and tabulated results for the pertinent Feynman integrals, mostly involving massless integrals. And the majority of the results given are only for the divergent parts of them and the complete closed form (or with the finite parts) of these are not known. In this manuscript, we use the technique of negative dimensional integration (NDIM) for the simplest of the one-loop massive integrals as a working bench for massive Feynman integrals in the light-front and show that novel results for the finite parts not known before are obtained. 

\end{abstract}



\maketitle

\section{Introduction}

\vspace{.5cm}

Our present understanding of elementary particles and their interactions is based on the principles built from quantum field theory. More specifically, gauge fields within the field theoretical construct in which gauge fields are a class of quantum fields where there exists a group of transformations of the field variables under which the basic physical properties remain unchanged, e.g., invariance of action leading to field equations remaining the same.  Those field variable transformations are called gauge transformations and
the principle of gauge invariance is foundational to our current understanding of particles and their interactions. A related but very often misunderstood concept is the question of gauge independent quantities. This has to do with the fact that relevant physical quantities of interest in any phenomenon must be gauge-independent, i.e., whatever the gauge choice we make use of, the resulting measurable quantity must be independent of the choice made. In other words, calculations of physical quantities performed in any gauge must give the same result.

Gauge invariance entails the necessity of gauge fixing procedures, allowing several different types of gauge choices to do the fixing. Among many, the physical light-cone gauge is the one that is defined through an external, constant, light-like four-vector $n^\mu$, $n^2 = 0$, but it is now well-understood that this single vector is not sufficient to span the whole four-dimensional space-time \cite{Leibbrandt1}.  In other words, it means that the gauge freedom is not totally removed. Residual gauge freedom remains to be removed by the introduction of the dual four-vector $n^{*\mu}$, $n^{*2} = 0$. This can done by hand using the {\it ad-hoc} Mandelstam-Leibbrandt (ML) prescription \cite{ML, Leibbrandt1a} in the calculations, whereas in the negative dimensional integration method (NDIM) \cite{Halliday-Ricotta} approach we use here, this is done naturally as a consequence of the general structure of the light-cone integral, defined over four-dimensional Minkowski space-time. The use of ML prescription can be seen in a variety of light-cone papers published in the literature, attesting that it suits and works for handling perturbative Feynman integral calculations \cite{LCML}. Pertinent integrals in the light-cone are evaluated and their pole parts are even tabulated in various places, but conspicuously, they are limited to massless integrals and only very few massive ones \cite{Basseto, Leibbrandt2}. 

The negative-dimensional integration approach  does not require the use of any prescriptions \cite{Suzuki} and provides physically acceptable results, i.e., causality preserving ones. The calculation we will present here is the very first test for massive light-cone integrals at the one-loop order without invoking the ML-prescription. The NDIM technique demonstrates that integration over components and partial fractioning tricks can be completely avoided, as well as parametric integrals. The important point to note is that the dual light-like four-vector $n^*$ is necessary in order to span the whole four-dimensional space, when defining the gauge proper. In the course of calculations, integrals are dimensionally regularized into a $D$-dimensional space-time.
Now we apply such technique to the simplest massive one-loop integral. The reason for this is two-fold: first, the massless one-loop tadpole type integral in the light-cone is known to be relevant for the perturbative calculations \cite{Capper}, contrary to what we have in the covariant case, where integrals of this type can be consistently set to zero. Also, the result for the tadpole integral in the light-cone we know in literature is restricted to the divergent part only; it is not calculated explicitly, with its finite parts. In other words, only the pole structure of it is known, showing that it does not contain any pathological features. Secondly, because in the NDIM approach we use, a whole class of integrals is calculated simultaneously, the integral (\ref{Integral}) being just a particular case of ours, and the results show a set of equivalent results at the same time, the other result in the set not known previously.

The study of relativistic dynamics was pioneered by Dirac \cite{Dirac}, who considered three possibilities for such description, among which the front form is a well-defined alternative for describing relativistic fields, characterized by the time evolution along the plane $x^+=0$. The variables $x^+ = (x^0+x^3)/\sqrt{2}$ and $x^-=(x^0-x^3)/\sqrt{2}$ are traditionally understood as the light-front ``time'' and longitudinal space variables, respectively; the transverse components remaining as in the usual Minkowski space-time, ${\bf x}^\perp=(x,y) \equiv (x^1,x^2)$. Their conjugate momenta are also similarly defined as $k^+=(k^0+k^3)/\sqrt{2}$, $k^-=(k^0-k^3)/\sqrt{2}$ and ${\bf k}^\perp = (k^1,k^2)$, being understood as the longitudinal momentum, light-front ``energy'' and transverse momentum respectively.

For a massive particle, the on-shell condition $k^2=m^2$ leads to
\be
k^- = \frac{{\bf k}_\perp^2+m^2}{2k^+}. \label{dispersion}
\ee

The dispersion relation in Eq.(\ref{dispersion}) is quite remarkable for the following reasons: (a) Even though it is a relativistic energy-momentum relation, it is a linear relation, contrary to the usual quadratic one; (b) The dependence of the light-front ``energy'' $k^-$ with respect to the transverse momentum ${\bf k}^\perp$ is just like the non-relativistic relation. (c) There is a sign correlation between the longitudinal momentum $k^+$ and the energy $k^-$; for $k^+$ positive (negative), $k^-$ is also positive (negative). (d) The dependence of the energy $k^-$ on the momentum components ${\bf k}^\perp$ and $k^+$ is multiplicative and large energy can result from large ${\bf k}^\perp$ and/or small $k^+$. All these simple observations have dramatic consequences into the relativistic physical aspects of particle dynamics.

Our notation and conventions for light front coordinates are as follows:
\bea
\label{conventions}
\left[ \begin{array}{c}
x^{{+}}  \\
x^{{-}}
\end{array}
\right] = \frac{1}{\sqrt{2}}\left[ \begin{array}{cc} 1 & 1 \\
1 & -1 \end{array} \right]\left[\begin{array}{c} x^0 \\ x^3  \end{array}\right],
\eea 
and therefore
\bea
\left[ \begin{array}{c}
x^0  \\
x^3
\end{array}
\right] = \frac{1}{\sqrt{2}}\left[ \begin{array}{cc}1& 1 \\
1 & -1 \end{array} \right]\left[\begin{array}{c} x^{+} \\ x^{-}  \end{array}\right].
\eea 

We have therefore the Minkowski space-time metric $g^{{\mu}{\nu}} = (+,\,-,\,1,\,2)$ given by 
\bea
g^{{\mu}{\nu}} =
\left[ \begin{array}{rrrr}
0 & 1 & 0 & 0  \\
1 & 0 & 0 & 0 \\
0 & 0 & -1 & 0 \\
0 & 0 & 0 & -1 
\end{array} \right] = g_{{\mu}{\nu}}.
\eea

This means that the covariant and contravariant indices of a given vector are related by
\bea
\label{definitions}
a_{+}  =   a^{-} & ; & a^{+}  =  a_{-} \nonumber \\
a_{{j}}  =  -a^{{j}} & , &  \mbox{(${j}=1,\,2$)}.
\eea

Then the scalar product of any two vectors becomes
\bea
a_{{\mu}} b^{{\mu}} & = & a_{+}b^{+}+a_{-}b^{-}+{\bf a}_{\perp}\cdot {\bf b}^{\perp}\nonumber\\
&=& a^-b^++a^+b^--{\bf a}^\perp \cdot {\bf b}^\perp,
\eea
where we use the convenient shorthand ${\bf a}_\perp =(a_1,a_2)$ and ${\bf a}^\perp =(-a^1,-a^2)$.

Using these conventions, our light-cone defining external vector $n_\mu = (1,0,0,1)/\sqrt{2}$ is now written as  $n_\mu = (n_+,n_-,n_1,n_2)=(1,0,0,0)$ while $n^\mu = (n^+,n^-,n^1,n^2)=(0,1,0,0)$.

Since the defining light-like external vector is arbitrary, we may equally choose the corresponding dual vector $n^{*}_\mu = (n^*_0,n^*_1,n^*_2,n^*_3)=(1,0,0,-1)/\sqrt{2}$, or equivalently, $n^*_\mu=(n^*_+,n^*_-,n^*_1,n^*_2)=(0,1,0,0)$. In such a choice of vectors we have $n^2=n^{*2}=0$ and $n\cdot n^*=1$.

\section{Massive One-loop Light-cone Feynman Integral}

Let us consider the simplest basic one-loop light-cone integral with massive propagator, in a generic $D$-dimensional space-time (omitting the usual $(2\pi)^D$ factor in the denominator of the integrand for convenience):
\bea
\label{Integral}
T_{D}(p,m^2)  & = &   \int  \frac{d^{D}\!k}{\left[ (k-p)^2-m^2+i\varepsilon\right]k\cdot n} \nonumber\\
             & = &   \int  \frac{d^{D}\!k}{\left[ (k-p)^2-m^2+i\varepsilon\right]k^+}.
\eea

Strictly speaking, we should have written $T_{D}(p,m^2, n, n^*)$, however, considering that $p^\mu$ can be expressed as $p^\mu = (p^+, p^1, p^2,p^-)\equiv(p\cdot n, p^1, p^2,p\cdot n^*) $, we drop the explicit dependence on $n$ and $n^*$.  The divergent part of (\ref{Integral}) is known for a long time and can be seen, for example, in the Appendices of Bassetto's and Leibbrandt's book \cite{Basseto}, \cite{Leibbrandt2} and it reads, with the use of ML prescription,
\bea
\label{ML_massive}
T_{D}(p,m^2) & = & \frac{2p\cdot n^*}{n\cdot n^*}I_{\rm div.} +F\nonumber\\
            & = & 2p^-I_{\rm div.} +F, \quad {\rm using} \:\,p\cdot n^* \equiv p^- \:\, {\rm and} \:\,n\cdot n^* =1. 
\eea
In the above equation, $F$ is the finite part, for which no specific value is given, and $I_{\rm div.}$ is the divergent part of the following integral which is dependent on the dimensionality $D$ of the space-time:
\bea
I_{\rm div.} & \equiv  & {\rm divergent\: part \:of} \int \frac{d^{D}\!k}{(k^2+i\varepsilon)[(k-p)^2+i\varepsilon]}\nonumber \\
                  & = & \left \{ \begin{array}{ll} \displaystyle\frac{\pi^2}{2-D/2},\quad {\rm Euclidean \:space.}\\ \displaystyle\frac{i\pi^2}{2-D/2}, \quad {\rm Minkowski\:space.}\end{array} \right.
\eea

Note that in (\ref{ML_massive}) we have the emergence of the dual vector $n^*$ in the result. It can be seen that as $D \to 4$ this term $I_{\rm div.}$ clearly diverges. 

We are going to evaluate an integral which is more general than (\ref{Integral}), where the denominator factors are raised to generic powers, using the technique of negative dimensional integration, without the need of any prescription to handle the $(k\cdot n)^{-j}\equiv (k^+)^{-j}$ pole, $j \subset {\mathbb N}$.

\subsection{The NDIM technique in Euclidean space}

For the purposes of applying the NDIM technique, we first introduce the general structure of the light-cone integral to be evaluated, namely,
\bea
\label{NDIM_massive}
T^\prime_{D}(p,m^2,i,j, l) & = & \int  \frac{d^{D}\!k}{\left[ (k-p)^2-m^2+i\varepsilon\right]^{i}}\frac{(k\cdot n^*)^{l}}{(k\cdot n)^{j}}.
\eea

Then, our original massive light-cone integral (\ref{Integral}) is a particular case of our general structure integral $T^\prime_D(p,m^2,i,j, l)$ for the specific values of exponents $(i,j,l)=(1,1,0)$.

To implement the NDIM technique for the sought integral, let us introduce the generating functional Gaussian integral that is pertinent to our calculation:
\bea
\label{Gaussian_generating}
G_D(p,m^2) & = & \int  d^{D}\!k \;e^{-\alpha[(k-p)^2-m^2]-\beta k\cdot n-\gamma k\cdot n^*} \\
                    & = & e^{\alpha(m^2-p^2)}\int d^D \!k\; e^{-\alpha\left(k^2-2k\cdot p +\frac{\beta}{\alpha}k\cdot n+\frac{\gamma}{\alpha}k\cdot n^*\right)}.
\eea

Completing the square in the argument of the exponential in the integrand, we can evaluate it, resulting in
\bea
G_D(p,m^2) & = & \frac{\pi^{D/2}}{\alpha^{D/2}}\,e^{\alpha m^2 - \beta p\cdot n - \gamma p\cdot n^* +\frac{\beta\gamma}{2\alpha}n\cdot n^*}
\eea

Next, we are going to expand the exponential function in the result above in power series to get
\bea
\label{Gaussian_result}
G_D(p,m^2) & = & \pi^{D/2}\!\!\!\sum_{a,b,c,d=0}^{\infty} (-1)^{b+c} \alpha^ {a-d-D/2}\beta^{b+d}\gamma^{c+d} \frac{(m^2)^a}{a!}\frac{(p\cdot n)^b}{b!}\frac{(p\cdot n^*)^c}{c!}\frac{(n\cdot n^*)^d}{2^d d!}.
\eea

On the other hand, expanding in power series the original Gaussian integral (\ref{Gaussian_generating}), we have
\bea
\label{Gaussian_NDIM}
G_D(p,m^2) & = & \int  d^{D}\!k \;e^{-\alpha[(k-p)^2-m^2]-\beta k\cdot n-\gamma k\cdot n^*} \nonumber\\
                    & = & \sum_{i,j,l=0}^{\infty} (-1)^{i+j+l}\frac{\alpha^i \beta^j \gamma^ l}{i!j!l!}\int d^D\!k\left[(k-p)^2-m^2\right]^i(k\cdot n)^j(k\cdot n^*)^l\nonumber\\
                                        & \equiv & \sum_{i,j,l=0}^{\infty} (-1)^{i+j+l}\frac{\alpha^i \beta^j \gamma^ l}{i!j!l!}I^{\rm s}_{\rm NDIM}(i,j,l),
\eea
where we have introduced the negative dimensional integral definition
\bea
\label{I_NDIM}
I^{\rm s}_{\rm NDIM}(i,j,l) & = & \int d^D\!k\left[(k-p)^2-m^2\right]^i(k\cdot n)^j(k\cdot n^*)^l.
\eea

The upper index `s' labels the kind of function dependence that the NDIM solution allows. Comparing (\ref{I_NDIM}) with (\ref{NDIM_massive}) we observe that the general structure light-cone integral in (\ref{NDIM_massive}) is reproduced by (\ref{I_NDIM}) when $i\to -i, j\to-j, l\to l$. This means that we need to make an analytic continuation for the pertinent indices $(i,j) \xrightarrow{\rm AC} (-i, -j)$ to allow for negative values of exponents $i$ and $j$, while $l$ remains unchanged in (\ref{I_NDIM}). After this analytic continuation process we then have the final result for the integral in (\ref{NDIM_massive}) as
\bea
T^\prime_D(p,m^2,i,j, l) = \sum_{\rm s} I^{\rm s(AC)}_{\rm NDIM}(i,j,l).
\eea

Next, we see that both (\ref{Gaussian_result}) and (\ref{Gaussian_NDIM}) are expressed in terms of series expansion in $\alpha, \beta, {\rm and}\, \gamma$. So these two series to be equal requires that term by term they must be equal. However, in the former we have four summation indices and in the latter only three summation indices. Identity between the two equations (\ref{Gaussian_result}) and (\ref{Gaussian_NDIM}) requires that:
\bea
i & = & a-d-D/2 \nonumber \\
j & = & b+d \\
l & = & c+d \nonumber
\eea

We want to solve the set $(a,b,c,d)$ in terms of $(i,j,l)$ set. Since there is one summation index more than the number of equalities, we can only solve the system of three equations with four unknowns in terms of one of the unknowns. Because in this case we have four unknowns, there are four possible resolutions for the system. These will be expressed in terms of a single sum whose index is one of the unknowns. Also, these resolutions will define the category of functional dependence `s' to which a given resolution will lead into. So let us detail this procedure in the following.

Solutions in $a$ and $d$ indices lead respectively to the following resolutions for the system:
\bea
\label{I_NDIM_1}
I^{[1]\rm s=z}_{\rm NDIM}(i,j,l) & = & (-\pi)^{D/2} i!j!l!\left(-\frac{m^2}{z}\right)^{i+D/2}(p\cdot n)^j (p\cdot n^*)^l\nonumber\\
&&\times \sum_{a=0}^{\infty}\frac{1}{(i+j+D/2-a)!(i+l+D/2-a)!(-i-D/2+a)!}\frac{z^a}{a!},
\eea
and 
\bea
\label{I_NDIM_2}
I^{[2]\rm s=z}_{\rm NDIM}(i,j,l) & = & (-\pi)^{D/2} i!j!l!\left(-m^2\right)^{i+D/2}(p\cdot n)^j (p\cdot n^*)^l\nonumber\\
&&\times \sum_{d=0}^{\infty}\frac{1}{(i+D/2+d)!(j-d)!(l-d)!}\frac{z^d}{d!},
\eea
where the variable $z$ is given by the following mass-momentum ratio:
\bea
z \equiv \frac{m^2 n\!\cdot \!n^*}{2p\!\cdot \!n  p\!\cdot\! n^*} = \frac{m^2}{2p^+p^-} .
\eea

Solutions in $b$ and $c$ indices lead respectively to the following resolutions for the system:
\bea
\label{I_NDIM_3}
I^{[3]{\rm s}=w}_{\rm NDIM}(i,j,l) & = & (-\pi)^{D/2} i!j!l!\left(-m^2\right)^{i+j+D/2}\left(-\frac{n\!\cdot\!n^*}{2p\!\cdot\! n^*}\right)^j (p\cdot n^*)^l\nonumber\\
&&\times \sum_{b=0}^{\infty}\frac{1}{(i+j+D/2-b)!(-j+l+b)!(j-b)!}\frac{w^b}{b!},
\eea
and 
\bea
\label{I_NDIM_4}
I^{[4]{\rm s}=w}_{\rm NDIM}(i,j,l) & = & (-\pi)^{D/2} i!j!l!\left(-m^2\right)^{i+l+D/2}(p\cdot n)^j \left(-\frac{n\!\cdot\! n^*}{2p\!\cdot\!n}\right)^l\nonumber\\
&&\times \sum_{c=0}^{\infty}\frac{1}{(i+l+D/2-c)!(j-l+c)!(l-c)!}\frac{w^c}{c!},
\eea
where the variable $w$ is given by the following mass-momentum ratio:
\bea
w \equiv \frac{1}{z} = \frac{2p\!\cdot \!n  p\!\cdot\! n^*}{m^2 n\!\cdot \!n^*} =\frac{2p^+p^-}{m^2} .
\eea

We now use the Euler's gamma function representation for the factorials and introduce the Pochhammer's symbol notation:
\bea
i! & = & \Gamma (1+i) \\
\Gamma (i-a) & = &  \frac{\Gamma(i-a)}{\Gamma(i)}\Gamma (i) \equiv (i)_{-a}\Gamma (i) ,
\eea
where in the last line we have defined the Pochhammer's symbol notation for the ratio of gamma functions:
\bea
(x)_y \equiv  \frac{\Gamma (x+y) }{\Gamma (x)}.
\eea

Within the summation sign, we will encounter Pochhammer's symbol with negative values for the summation index, so we need to rewrite them using the well-known analytic continuation property from negative to positive values. For example, 
\bea 
\label{Poch}
(1+n)_{-a} = \frac{(-1)^a}{(1-(1+n))_a}=\frac{(-1)^a}{(-n)_a}.
\eea

Then, our resolutions can be written as 
\bea
I^{[1]\rm s=z}_{\rm NDIM}(i,j,l) & = & (-\pi)^{D/2}\left(-\frac{m^2}{z}\right)^{i+D/2}\frac{(p\cdot n)^j (p\cdot n^*)^l}{(1+i)_{-2i-D/2}(1+j)_{i+D/2}(1+l)_{i+D/2}}\nonumber\\
&&\times \sum_{a=0}^{\infty}\frac{(-i\!-\!j\!-\!D/2)_a (-i\!-\!l\!-\!D/2)_a}{(1\!-\!i\!-\!D/2)_a}\frac{z^a}{a!},\\
I^{[2]\rm s=z}_{\rm NDIM}(i,j,l) & = & (-\pi)^{D/2} \left(-m^2\right)^{i+D/2}(p\cdot n)^j (p\cdot n^*)^l\frac{1}{(1+i)_{D/2}}\nonumber\\
&&\times \sum_{d=0}^{\infty}\frac{(-j)_d(-l)_d}{(1\!+\!i+\!D/2)_d}\frac{z^d}{d!},
\eea

The single left over sums are the well-known hypergeometric functions of one variable, so we have
\bea
I^{[1]\rm s=z}_{\rm NDIM}(i,j,l) & = & (-\pi)^{D/2}\left(-\frac{m^2}{z}\right)^{i+D/2}\frac{(p\cdot n)^j (p\cdot n^*)^l}{(1+i)_{-2i-D/2}(1+j)_{i+D/2}(1+l)_{i+D/2}}\nonumber\\
&&\times {}_2F_1\left(\left. -i\!-\!j\!-\!D/2, \; -i\!-\!l\!-\!D/2;\;1\!-\!i\!-\!D/2\,\right| z \right)\\
I^{[2]\rm s=z}_{\rm NDIM}(i,j,l) & = & (-\pi)^{D/2} \left(-m^2\right)^{i+D/2}(p\cdot n)^j (p\cdot n^*)^l\frac{1}{(1+i)_{D/2}}\nonumber\\
&&\times {}_2F_1\left(\left. \!-j,\,-l;\,1\!+\!i\!+\!D/2\,\right|z\right),
\eea

Final step is to analytic continue the coefficient Pochhammer's symbols allowing for negative values of $i$ and $j$, using (\ref{Poch}) so that the solution for the integral in the $z$ parameter is
\bea
\label{Integral_z}
T^\prime_D(z) & = & \sum_{\rm s} I^{\rm s=z(AC)}_{\rm NDIM}(i,j,l)\nonumber\\
                        & = &  \pi^{D/2}(-m^2)^{i+D/2}(p\cdot n)^{j}(p\cdot n^*)^l\left \{{\cal F}_{1}\left(-i\!-\!j\!-\!D/2,\,-i\!-\!l\!-\!D/2;1\!-i\!-\!\left.D/2\right|z\right)\right.\nonumber\\
                        && +\left.{\cal F}_{2}\left(\!-j,\!-l;1\!+i\!+\!\left.D/2\right|z\right)\right\},
\eea
in which we have introduced the following definitions:
\bea
{\cal F}_{1}(\cdots \,|z) & \equiv & \frac{\Gamma(i\!+\!D/2)\Gamma(-i\!-\!j\!-\!D/2)\Gamma(1\!+\!l)}{\Gamma(-i)\Gamma(-j)\Gamma(1\!+\!i\!+\!l\!+\!D/2)}\nonumber\\
&&\times (-z)^{-i-D/2}\left.{}_2F_{1}\left(-i\!-\!j\!-\!D/2,\,-i\!-\!l\!-\!D/2;1\!-i\!-\!\left.D/2\right|z\right)\right.,\\
{\cal F}_{2}(\cdots \,|z) & \equiv & \frac{\Gamma(-i\!-\!D/2)}{\Gamma(-i)}\left.{}_2F_{1}\left(\!-j,\!-l;1\!+i\!+\!\left.D/2\right|z\right)\right..
\eea

Similarly, we obtain for the other set of resolutions, 
\bea
\label{Integral_w}
T^\prime_D(w) & = & \sum_{\rm s} I^{{\rm s}=w{\rm(AC)}}_{\rm NDIM}(i,j,l)\nonumber\\
                        & = &  \pi^{D/2}(-m^2)^{i+D/2}(p\cdot n)^{j}(p\cdot n^*)^l\left \{{\cal F}_{3}\left(-j,\,-i\!-\!j\!-\!\left.D/2;1\!-j\!+\!l\right|w\right)\right.\nonumber\\
                        && +\left.{\cal F}_{4}\left(\!-l,\,-i\!-\!l\!-\!\left.D/2;1\!+j\!-\!l\right|w\right)\right\},
\eea
with
\bea
{\cal F}_{3}(\cdots \,|w) & \equiv & \frac{\Gamma(-i\!-\!j\!-\!D/2)\Gamma(1\!+\!l)}{\Gamma(-i)\Gamma(1\!-\!j\!+l)}(-w)^{-j}\left.{}_2F_{1}\left(-j,-i\!-\!j\!-\!\left.D/2;1\!-j\!+l\right|w\right)\right., \label{w1}\\
{\cal F}_{4}(\cdots \,|w) & \equiv & \frac{\Gamma(-i\!-\!l\!-\!D/2)\Gamma(-j\!+\!l)}{\Gamma(-i)\Gamma(-j)}\left. (w)^{-l}\:{}_2F_{1}\left(-l,\,-i\!-\!l\!-\!\left.D/2;1\!+j\!-\!l\right|w\right)\right.. \label{w2}
\eea

So far then, we have two possible sets as candidates to be solutions for our integral (\ref{NDIM_massive}) using the NDIM technique, one with momentum configuration $z$ and the other one, $w$. However, in the second set of solutions, (\ref{Integral_w}), the two Gauss hypergeometric functions in (\ref{w1}) and (\ref{w2}) are {\it not} linearly independent of each other, since the third parameter, $1-j+l$ (or equivalently, $1+j-l$), is an integer number for $j, l$ integers \cite{GR}. This linear dependency can be lifted by using the following Gauss' hypergeometric function relation \cite{GR}
\bea
{}_2F_{1}\left(a,\,b;\,c\left |x\right.\right) &=& \frac{\Gamma(c)\Gamma(b-a)}{\Gamma(b)\Gamma(c-a)}(-x)^{-a}{}_2F_{1}\left(a,\,a+1-c\,;a+1-b\left|\right. x^{-1}\right) \nonumber\\
                                                              &+& \frac{\Gamma(c)\Gamma(a-b)}{\Gamma(a)\Gamma(c-b)}(-x)^{-b}{}_2F_{1}\left(b,\,b+1-c\,;b+1-a\left|\right. x^{-1}\right). \label{HGFCID}
\eea

Using (\ref{HGFCID}) in the hypergeometric function of (\ref{w1}) we get
\bea
{}_2F_{1}\left(-j,-i-j-D/2;1-j+l\left |w\right.\right) &=& \frac{(1+l)_{-j}}{(-i-D/2)_{-j}}\left(-w\right)^{j}{}_2F_{1}\left(-j,-l;1+i+D/2\left|\right. z\right) \nonumber\\
                                                              &+& \frac{(-j)_{1+l}}{(i+D/2)_{1+l}}(-w)^{i+j+D/2}{}_2F_{1}\left(\alpha,\beta;\gamma\left|\right. z\right), \label{ID}
\eea
where ${}_2F_{1}\left(\alpha,\beta;\gamma\left|\right. z\right)\equiv {}_2F_{1}\left(-i-j-D/2,-i-l-D/2;1-i-D/2\left|\right. z\right)$ in the last term in (\ref{ID}). Substituting (\ref{ID}) in the first term of (\ref{Integral_w}), we obtain exactly (\ref{Integral_z}). This means that the second term appearing in (\ref{Integral_w}), proportional to ${\cal F}_4(-l,-i-l-D/2;1+j-l|w)$ is superfluous, since, as mentioned before, does not constitute a linearly independent solution in relation to ${\cal F}_3(-j,-i-j-D/2;1-j+l|w)$.

We may ask whether the second term (\ref{w2}) in (\ref{Integral_w}) can be considered the solution in the variable $w$ while the first term (\ref{w1}) there be considered superfluous. The answer is yes, indeed, since as we mentioned before, the two terms in (\ref{Integral_w}) are not linearly independent from each other. We can show that using the same identity (\ref{HGFCID}) in the hypergeometric function (\ref{w2}) leads exactly to the same solution in the variable $z$ given in (\ref{Integral_z}), although the algebra of Pochhammer's symbols in this case becomes more involved in proving this.

Therefore, with the NDIM technique applied for the massive Feynman integral (\ref{NDIM_massive}) with general powers for the propagators, we have two possible, distinct, sets of solutions for our integral (\ref{NDIM_massive}), namely, 
\bea
T^\prime_D(z)  & = & {\bf C}(p,m^2)\left \{{\cal F}_{1}\left(-i\!-\!j\!-\!D/2,\,-i\!-\!l\!-\!D/2;1\!-i\!-\!\left.D/2\right|z\right)\right.\nonumber\\
                        && +\left.{\cal F}_{2}\left(\!-j,\!-l;1\!+i\!+\!\left.D/2\right|z\right)\right\}, \label{scalarz}
\eea
and
\bea
T^\prime_D(w)  & = & {\bf C}(p,m^2)\left \{{\cal F}_{3}\left(-j,-\!i\!-j\!-\!D/2;\,1-\!j\!+\!l\right|w\right), \label{scalarw}
\eea
where we have defined ${\bf C}(p,m^2) \equiv \pi^{D/2}(-m^2)^{i+D/2}(p\cdot n)^{j}(p\cdot n^*)^l$.  These two sets of answers were not known previously in such a closed form. In the appendix we present a detailed calculation showing that for the special case of these results with particular values of the exponents, namely, $i=j=-1$ and $l=0$. In this particular case,  equation (\ref{NDIM_massive}) becomes just the integral in (\ref{Integral}), and we show its concordance with result (\ref{ML_massive}).

\subsection{The NDIM technique with tensorial structure in Euclidean space}

Next, let us consider the simplest basic one-loop light-cone integral with massive propagator and with tensorial structure in the numerator, in a generic $D$-dimensional space-time (omitting the usual $(2\pi)^D$ factor in the denominator of the integrand for convenience):
\bea
\label{tensor}
T_{D}^{\mu}(p,m^2)  & = &   \int  \frac{d^{D}\!k \:\:\:k^{\mu}}{\left[ (k-p)^2-m^2+i\varepsilon\right]k\cdot n} \nonumber\\
             & = &   \int  \frac{d^{D}\!k \:\:\:k^{\mu}}{\left[ (k-p)^2-m^2+i\varepsilon\right]k^+}.
\eea

As before, we introduce the general structure (for convenience we introduce a factor 2 in both numerator $2k\cdot n^*$ and denominator $2k\cdot n$)
\bea
\label{NDIM_tensor}
T^{\prime \mu}_{D}(p,m^2,i,j, l) & = & \int  \frac{d^{D}\!k\:\:\: k^{\mu}}{\left[ (k-p)^2-m^2+i\varepsilon\right]^{i}}\frac{(2k\cdot n^*)^{l}}{(2k\cdot n)^{j}},
\eea
with its corresponding generating functional Gaussian integral 
\bea
\label{Gaussian_tensor}
G_D^{\mu}(p,m^2) & = & \int d^D \!k\; k^\mu\:e^{-\alpha\left[(k-p)^2-m^2\right]-2\beta k\cdot n-2\gamma k\cdot n^*}.
\eea

The above can be rewritten using the identity
\be
k^\mu e^{2\alpha k\cdot p} \equiv \frac{1}{2\alpha}\frac{\partial}{\partial p_\mu} e^{2\alpha k\cdot p}, 
\ee
yielding 
\bea
\label{Gaussian_tensor1}
G_D^{\mu}(p,m^2) & = & e^{\alpha(m^2-p^2)}\int d^D \!k\; k^\mu\:e^{-\alpha\left(k^2-2k\cdot p +\frac{2\beta}{\alpha}k\cdot n+\frac{2\gamma}{\alpha}k\cdot n^*\right)}\nonumber\\        
                               & = & e^{\alpha(m^2-p^2)}\left(\frac{1}{2\alpha}\frac{\partial}{\partial p_\mu}\right)\int d^D \!k\:e^{-\alpha\left(k^2-2k\cdot p +\frac{2\beta}{\alpha}k\cdot n+\frac{2\gamma}{\alpha}k\cdot n^*\right)}.
\eea

Performing the Gaussian momentum integration, we get
\bea
\label{Gaussian_tensor2}
G_D^{\mu}(p,m^2) & = & \pi^{D/2}\left\{\frac{1}{\alpha^D}p^\mu-\frac{\beta}{\alpha^{D+1}}n^\mu-\frac{\gamma}{\alpha^{D+1}}n^{*\mu}\right\} e^{\alpha m^2-2\beta p\cdot n-2\gamma p\cdot n^* +\frac{2\beta\gamma}{\alpha}n\cdot n^*}.
\eea

Applying the partial differential operator on the result and expanding the exponential function in series, we get
\bea
\label{Gaussian_expand}
G_D^{\mu}(p,m^2) & = & \pi^{D/2} \left\{\frac{1}{\alpha^D}p^\mu-\frac{\beta}{\alpha^{D+1}}n^\mu-\frac{\gamma}{\alpha^{D+1}}n^{*\mu}\right\} \nonumber\\
                              &\times& \sum_{a,b,c,d=0}^{\infty}(-1)^{b+c}\alpha^{a-d}\beta^{b+d}\gamma^{c+d}\frac{(m^2)^a}{a!}\frac{(2 p\cdot n)^b}{b!}\frac{(2p\cdot n^*)^c}{c!}\frac{(2 n\cdot n^*)^d}{d!}.
\eea

Different from the previous example of the scalar tadpole integral, now we have a splitting into three relevant terms,
\be
G_D^{\mu}(p,m^2) = \pi^{D/2} \left\{p^\mu G_D(p)-n^\mu G_D(n)-n^{*\mu}G_D(n^{*})\right\},
\ee
with
\bea
G_D(p)  \!\!\!& = \!\!\!& \sum_{a,b,c,d=0}^{\infty}(-1)^{b+c}\alpha^{a-d-D/2}\beta^{b+d}\gamma^{c+d}\frac{(m^2)^a}{a!}\frac{(2 p\cdot n)^b}{b!}\frac{(2p\cdot n^*)^c}{c!}\frac{(2 n\cdot n^*)^d}{d!},\\
G_D(n)  \!\!& = & \!\!\! \sum_{a,b,c,d=0}^{\infty}(-1)^{b+c}\alpha^{a-d-D/2-1}\beta^{b+d+1}\gamma^{c+d}\frac{(m^2)^a}{a!}\frac{(2 p\cdot n)^b}{b!}\frac{(2p\cdot n^*)^c}{c!}\frac{(2 n\cdot n^*)^d}{d!},\\
G_D(n^*)  \!\!& = & \!\!\! \sum_{a,b,c,d=0}^{\infty}(-1)^{b+c}\alpha^{a-d-D/2-1}\beta^{b+d}\gamma^{c+d+1}\frac{(m^2)^a}{a!}\frac{(2 p\cdot n)^b}{b!}\frac{(2p\cdot n^*)^c}{c!}\frac{(2 n\cdot n^*)^d}{d!}.
\eea

The original Gaussian integral (\ref{Gaussian_tensor}) on the other hand has the following series expansion
\bea
\label{GT}
G_D^{\mu}(p,m^2) & = & \sum_{i,j,l=0}^{\infty}(-1)^{i+j+l}\frac{\alpha^i \beta^j \gamma^l}{i!j!l!}\int d^D \!k\; k^\mu\:\left[(k-p)^2-m^2\right]^i (2k\cdot n)^j (2k\cdot n^*)^l.
\eea

Defining 
\be
I_{\rm NDIM}^\mu (i,j,l) = \int d^D \!k\; k^\mu\:\left[(k-p)^2-m^2\right]^i (2k\cdot n)^j (2k\cdot n^*)^l,
\ee
we have, equating the series,
\be
I_{\rm NDIM}^\mu (i,j,l) = \pi^{D/2} \left\{p^\mu I_{\rm NDIM}(p)-n^\mu I_{\rm NDIM}(n)-n^{*\mu}I_{\rm NDIM}(n^*) \right\}
\ee
where 
\bea
I_{\rm NDIM}(p) & = & (-1)^{-i-j-l}i!j!l! {\sum} ' \, (-1)^{b+c}\alpha^{a-d-D/2} \beta^{b+d} \gamma^{c+d}\nonumber \\
                          &\times & \frac{(m^2)^a}{a!}\frac{(2 p\cdot n)^b}{b!}\frac{(2p\cdot n^*)^c}{c!}\frac{(2 n\cdot n^*)^d}{d!},\\
I_{\rm NDIM}(n) & = & (-1)^{-i-j-l}i!j!l! {\sum} '' \, (-1)^{b+c}\alpha^{a-d-D/2-1} \beta^{b+d+1} \gamma^{c+d}\nonumber \\
                          &\times & \frac{(m^2)^a}{a!}\frac{(2 p\cdot n)^b}{b!}\frac{(2p\cdot n^*)^c}{c!}\frac{(2 n\cdot n^*)^d}{d!},\\
I_{\rm NDIM}(n^*) & = & (-1)^{-i-j-l}i!j!l! {\sum} ''' \, (-1)^{b+c}\alpha^{a-d-D/2-1} \beta^{b+d} \gamma^{c+d+1}\nonumber \\
                          &\times & \frac{(m^2)^a}{a!}\frac{(2 p\cdot n)^b}{b!}\frac{(2p\cdot n^*)^c}{c!}\frac{(2 n\cdot n^*)^d}{d!},                      
\eea
with $\sum '$, $\sum ''$ and $\sum''' $ indicating constrained sum indices as follows
\begin{center}
\begin{tabular}{ |c|c|c| } 
 \hline
 $\sum '$ & $\sum ''$ & $\sum '''$ \\ 
 \hline
 $i=a-d-D/2$ & $i=a-d-D/2-1$ & $i=a-d-D/2-1$ \\ 
 $j=b+d$ & $j=b+d+1$ & $j=b+d$ \\ 
 $l=c+d$ & $l=c+d$ & $l=c+d+1$ \\ 
 \hline
\end{tabular}
\end{center}

As before, since we have more indices in the set $\{a,b,c,d\}$ than in the set $\{i,j,l\}$, it follows that one index in the former set remains free, which leaves a summation over the values of that index. So, we have for each of the $\{I_{\rm NDIM}(p),I_{\rm NDIM}(n),I_{\rm NDIM}(n^*)\}$ a set of solutions pertaining to four summations left, namely, $\sum_a$, $\sum_b$, $\sum_c$, and $\sum_d$. Each of these then should be analytically continued to positive dimension and then we analyze the solutions.

First, for the $I_{\rm NDIM}(p)$ we have, after being analytically continued to positive dimensions, 
\bea
I_{D}^{\prime a}(p) & = & {\mathbb C}^a(p) {}_2F_1(-i-j-D/2,\,-i-l-D/2;\,1-i-D/2|z), \label{tensor a}\\
I_{D}^{\prime b}(p) & = & {\mathbb C}^b(p) {}_2F_1(-j,\,-i-j-D/2;\,1-j+l|w), \label{tensor b}\\
I_{D}^{\prime c}(p) & = & {\mathbb C}^c(p) {}_2F_1(-l,\,-i-l-D/2;\,1+j-l|w), \label{tensor c} \\
I_{D}^{\prime d}(p) & = & {\mathbb C}^d(p) {}_2F_1(-j,\,-l;\,1+i+D/2|z), \label{tensor d}
\eea
with the following table of coefficient factors:
\begin{center}
Table 1
\vspace{.2cm}

\begin{tabular}{|c|c|} 
 \hline
 ${\mathbb C}^a(p) =$ & $(2p\cdot n)^{i+j+D/2}(2p\cdot n^*)^{i+l+D/2}(2n\cdot n^*)^{-i-D/2}\frac{\Gamma(i+D/2)\Gamma(-i-j-D/2)\Gamma(1+l)}{\Gamma(-i)\Gamma(-j)\Gamma(1+i+l+D/2)}$  \\ 
 \hline
 ${\mathbb C}^b(p) = $ & $ (-m^2)^{i+j+D/2}(2p\cdot n^*)^{-j+l}(2n\cdot n^*)^{j}\frac{\Gamma(-i-j-D/2)\Gamma(1+l)}{\Gamma(-i)\Gamma(1-j+l)}  $  \\ 
 \hline
 ${\mathbb C}^c(p) = $ & $ (-m^2)^{i+l+D/2}(2p\cdot n)^{j-l}(-2n\cdot n^*)^{l}\frac{\Gamma(-i-l-D/2)\Gamma(-j+l)}{\Gamma(-i)\Gamma(-j)}$  \\ 
 \hline
 ${\mathbb C}^d(p) = $ & $(-m^2)^{i+D/2}(2p\cdot n)^{j}(2p\cdot n^*)^l \frac{\Gamma(-i-D/2)}{\Gamma(-i)}$  \\ 
 \hline
\end{tabular}
\end{center}

Again, solutions (\ref{tensor b}) and (\ref{tensor c}) are not linearly independent since the third parameter of the hypergeometric function $1-j+l$ (or, alternatively, $1+j-l$) is an integer number for integer $j$ and $l$. 

Analogous analysis can be done here as has been done for the scalar tadpole results and we arrive at the following result:
\bea
I^{\prime}_D(p, z)  & = & {\mathbb C}^a(p) {}_2F_{1}\left(-i-j-D/2,\,-i-l-D/2;1-i-\left.D/2\right|z\right)\nonumber\\
                                     & + & {\mathbb C}^d(p) {}_{2}F_1\left(\!-j,\!-l;1\!+i\!+\!\left.D/2\right|z\right), \label{pz}
\eea
and
\bea
I^{\prime}_D(p, w)  & = & {\mathbb C}^b(p) {}_2F_1\left(-j,-i-j-D/2;\,1-j+l\left|w\right.\right). \label{pw}
\eea

Next, we do similar analysis for the $I_{\rm NDIM}(n)$ and obtain
\bea
I_{D}^{\prime a}(n) & = & {\mathbb C}^a(n) {}_2F_1(-i-j-D/2,\,-i-l-D/2-1;\,-i-D/2|z), \label{tensor na}\\
I_{D}^{\prime a}(n) & = & {\mathbb C}^b(n) {}_2F_1(-j+1,\,-i-j-D/2;\,2-j+l|w), \label{tensor nb}\\
I_{D}^{\prime a}(n) & = & {\mathbb C}^c(n) {}_2F_1(-l,\,-i-l-D/2-1;\,j-l|w), \label{tensor nc} \\
I_{D}^{\prime a}(n) & = & {\mathbb C}^d(n) {}_2F_1(-j+1,\,-l;\,2+i+D/2|z), \label{tensor nd}
\eea
with the following table of coefficient factors:

\begin{center}
Table 2
\vspace{.2cm}

\begin{tabular}{|c|c|} 
 \hline
 ${\mathbb C}^a(n) =$ & $(2p\cdot n)^{i+j+D/2}(2p\cdot n^*)^{i+l+D/2+1}(2n\cdot n^*)^{-i-D/2-1}\frac{\Gamma(i+D/2+1)\Gamma(-i-j-D/2)\Gamma(1+l)}{\Gamma(-i)\Gamma(-j)\Gamma(2+i+l+D/2)}$  \\ 
 \hline
 ${\mathbb C}^b(n) = $ & $ (-m^2)^{i+j+D/2}(2p\cdot n^*)^{-j+l+1}(2n\cdot n^*)^{j-1}\frac{\Gamma(-i-j-D/2)\Gamma(-j+1)\Gamma(1+l)}{\Gamma(-i)\Gamma(-j)\Gamma(2-j+l)}  $  \\ 
 \hline
 ${\mathbb C}^c(n) = $ & $ (-m^2)^{i+l+D/2+1}(2p\cdot n)^{j-l-1}(-2n\cdot n^*)^{l}\frac{\Gamma(-i-l-D/2-1)\Gamma(-j+l+1)}{\Gamma(-i)\Gamma(-j)}$  \\ 
 \hline
 ${\mathbb C}^d(n) = $ & $(-m^2)^{i+D/2+1}(2p\cdot n)^{j-1}(2p\cdot n^*)^l \frac{\Gamma(-i-D/2-1)\Gamma(-j+1)}{\Gamma(-i)\Gamma(-j)}$  \\ 
 \hline
\end{tabular}
\end{center}

These lead to
\bea
I^{\prime}_D(n, z)  & = & {\mathbb C}^a(n) {}_2F_{1}\left(-i-j-D/2,\,-i-l-D/2-1;-i-\left.D/2\right|z\right)\nonumber\\
                                     & + & {\mathbb C}^d(n) {}_{2}F_1\left(\!-j+1,\!-l;2\!+i\!+\!\left.D/2\right|z\right), \label{nz}
\eea
and
\bea
I^{\prime}_D(n, w)  & = & {\mathbb C}^b(n) {}_2F_1\left(-j+1,-i-j-D/2;\,2-j+l\left|w\right.\right). \label{nw}
\eea

From the analysis of $I_{\rm NDIM}(n^*)$ we get
\bea
I_{D}^{\prime a}(n^*) & = & {\mathbb C}^a(n^*) {}_2F_1(-i-j-D/2-1,\,-i-l-D/2;\,-i-D/2|z), \label{tensor n*a}\\
I_{D}^{\prime b}(n^*) & = & {\mathbb C}^b(n^*) {}_2F_1(-j,\,-i-j-D/2-1;\,-j+l|w), \label{tensor n*b}\\
I_{D}^{\prime c}(n^*) & = & {\mathbb C}^c(n^*) {}_2F_1(-l+1,\,-i-l-D/2;\,2+j-l|w), \label{tensor n*c} \\
I_{D}^{\prime d}(n^*) & = & {\mathbb C}^d(n^*) {}_2F_1(-j,\,-l+1;\,2+i+D/2|z), \label{tensor n*d}
\eea
with the following table of coefficient factors:
\begin{center}
Table 3
\vspace{.2cm}

\begin{tabular}{|c|c|} 
 \hline
 ${\mathbb C}^a(n^*) =$ & $-(2p\cdot n)^{i+j+D/2+1}(2p\cdot n^*)^{i+l+D/2}(2n\cdot n^*)^{-i-D/2-1}\frac{\Gamma(i+D/2+1)\Gamma(-i-j-D/2-1)\Gamma(1+l)}{\Gamma(-i)\Gamma(-j)\Gamma(1+i+l+D/2)}$  \\ 
 \hline
 ${\mathbb C}^b(n^*) = $ & $ -(-m^2)^{i+j+D/2+1}(2p\cdot n^*)^{-j+l-1}(2n\cdot n^*)^{j}\frac{\Gamma(-i-j-D/2-1)\Gamma(1+l)}{\Gamma(-i)\Gamma(-j+l)}  $  \\ 
 \hline
 ${\mathbb C}^c(n^*) = $ & $ -(-m^2)^{i+l+D/2}(2p\cdot n)^{j-l+1}(-2n\cdot n^*)^{l-1}\frac{\Gamma(-i-l-D/2)\Gamma(-j+l-1)\Gamma(1+l)}{\Gamma(-i)\Gamma(-j)\Gamma(l)}$  \\ 
 \hline
 ${\mathbb C}^d(n^*) = $ & $(-m^2)^{i+D/2+1}(2p\cdot n)^{j}(2p\cdot n^*)^{l-1} \frac{\Gamma(-i-D/2-1)\Gamma(-j+1)\Gamma(1+l)}{\Gamma(-i)\Gamma(l)}$  \\ 
 \hline
\end{tabular}
\end{center}

These result in:
\bea
I^{\prime}_D(n^*, z)  & = & {\mathbb C}^a(n^*) {}_2F_{1}\left(-i-j-D/2-1,\,-i-l-D/2;-i-\left.D/2\right|z\right)\nonumber\\
                                     & + & {\mathbb C}^d(n^*) {}_{2}F_1\left(-j,\,-l+1;2+i+\left.D/2\right|z\right), \label{n*z}
\eea
and
\bea
I^{\prime}_D(n^*, w)  & = & {\mathbb C}^b(n^*) {}_2F_1\left(-j,-i-j-D/2-1;\,-j+l\left|w\right.\right). \label{n*w}
\eea

Finally, collecting all the relevant results, we can write the results for the tensor structure integral (\ref{NDIM_tensor}):
\bea
T_D^{\prime \mu} & = & \pi^{D/2} \left\{ p^\mu I_D^\prime(p,z)- n^\mu I_D^\prime(n,z)-n^{*\mu} I_D(n^*,z) \right\} \label {z}\\
                             & = & \pi^{D/2} \left\{ p^\mu I_D^\prime(p,w)- n^\mu I_D^\prime(n,w)-n^{*\mu} I_D(n^*,w) \right\}. \label{w}
\eea

The first solution, (\ref{z}), in the above contains six hypergeometric functions in the variable $z=\frac{m^2 n\cdot n^*}{2p\cdot n p\cdot n^*}$. It reads: 
\bea
T_D^{\prime \mu}(z) & = & \pi^{D/2} p^\mu \left\{ {\mathbb C}^a(p) {}_2F_{1}\left(-i-j-D/2,\,-i-l-D/2;1-i-\left.D/2\right|z\right)\right.\nonumber\\
                                     &  & \hspace{1.1cm}+\left.{\mathbb C}^d(p) {}_{2}F_1\left(\!-j,\!-l;1\!+i\!+\!\left.D/2\right|z\right)\right\} \nonumber\\
                                     &-& \pi^{D/2} n^\mu \left\{{\mathbb C}^a(n) {}_2F_{1}\left(-i-j-D/2,\,-i-l-D/2-1;-i-\left.D/2\right|z\right)\right.\nonumber\\
                                     &  &\hspace{1.1cm}+\left. {\mathbb C}^d(n) {}_{2}F_1\left(\!-j+1,\!-l;2\!+i\!+\!\left.D/2\right|z\right)\right\} \nonumber\\
                                     &-& \pi^{D/2} n^{*\mu}\left\{{\mathbb C}^a(n^*) {}_2F_{1}\left(-i-j-D/2-1,\,-i-l-D/2;-i-\left.D/2\right|z\right)\right. \nonumber\\
                                     &  &\hspace{1.1cm}+ \left.{\mathbb C}^d(n^*) {}_{2}F_1\left(-j,\,-l+1;2+i+\left.D/2\right|z\right)\right\}. \label{answerz}
\eea

The second solution, (\ref{w}), contains only three hypergeometric functions in the inverse momentum configuration variable $w=z^{-1}$ and reads:
\bea
T_D^{\prime \mu}(w) & = & \pi^{D/2} \left\{ p^\mu {\mathbb C}^b(p) {}_2F_1\left(-j,-i-j-D/2;\,1-j+l\left|w\right.\right)\right.\nonumber \\
                             && \hspace{.75cm}- n^\mu {\mathbb C}^b(n) {}_2F_1\left(-j+1,-i-j-D/2;\,2-j+l\left|w\right.\right)\nonumber \\
                             &&  \hspace{.75cm}\left.-n^{*\mu} {\mathbb C}^b(n^*) {}_2F_1\left(-j,-i-j-D/2-1;\,-j+l\left|w\right.\right) \right\}.\label{answerw}
\eea

The coefficient factors ${\mathbb C}^{(a,d,b)}$ are tabulated in Tables 1, 2 and 3 and the two sets of solutions (\ref{answerz}) and (\ref{answerw}) are completely equivalent whose explicit expressions were unknown before.

\section{Conclusions}

We have then shown that using the NDIM technique, it was possible to evaluate the massive integral (\ref{Integral}) and also the integral with a tensorial structure (\ref{tensor}) without any prescription to treat the light-cone pole $(k\cdot n)^{-j}$ and the resolution of the integral was carried out by solving sets of systems of linear equations. Moreover, the solutions we get are more general, with generic exponents for denominators and are complete, in the sense that we have two equivalent sets of solutions, each one given in terms of a particular mass-momentum configuration parameter, either $z$ or $w$. These results are all novel results, since up to now, only the divergent part of the integral was widely known and used in the pertinent available literature.  

\section{Appendix}

\subsection{Scalar case}

In this appendix, we give a detailed computation of the divergent and finite parts of the results we have obtained for the special case of exponents $i=j=-1$ and $l=0$. Then integral (\ref{NDIM_massive}) is given either by equation (\ref{scalarz}) or equation (\ref{scalarw}). For the first, with momentum variable $z$, it reads
\bea
\label{Integral_z-particular}
T^\prime_D(-1,-1,0;z) & = &  \pi^{D/2}(-m^2)^{D/2-1}\frac{1}{(p\cdot n)}\left \{ (-z)^{1-D/2}\frac{\Gamma(\!D/2-1)\Gamma(2\!-\!D/2)}{\Gamma(\!D/2)}\right.\nonumber\\
&&\times \left.{}_2F_{1}\left(2\!-\!D/2,\,1\!-\!D/2;2\!-\left.D/2\right|z\right)\right.\nonumber\\
                        &&\hspace{4.3cm} +\left.\Gamma(1\!-\!D/2)\left.{}_2F_{1}\left(\!1,\!0;\!\left.D/2\right|z\right)\right.\right\}.
\eea

Since the Gauss hypergeometric functions are such that they have the following properties applicable to our case above \cite{GR}
\bea
\left.{}_2F_{1}\left(\alpha,\,\beta;\left.\alpha\right|z\right)\right. & = & \left( 1-z\right)^{-\beta} \\
{}_2F_{1}\left(\alpha,\left. \, 0;\,\gamma\right|z\right) & = & 1,
\eea
it gives us
\bea
\label{Tw}
T^\prime_D(-1,-1,0;z) & = & \pi^{D/2}\frac{(-m^2)^{D/2-1}}{(p\cdot n)}\left \{ (-z)^{1-D/2}\frac{\Gamma\left(\!D/2-1\right)\Gamma\left(2\!-\!D/2\right)}{\Gamma\left(\!D/2\right)}\left(1-z\right)^{D/2-1}\right.\nonumber\\
                        &&\hspace{3.2cm} +\left.\Gamma\left(1\!-\!D/2\right)\right\}.
\eea

Using the gamma function identity relation $x\Gamma(x)=\Gamma(1+x)$  \cite{GR} we may rewrite (\ref{Tw})as
\bea
T^\prime_D(-1,-1,0;z) & = & \pi^{2-\epsilon}\frac{(-m^2)^{1-\epsilon}}{(p\cdot n)}\left \{ \frac{\Gamma\left(1-\epsilon\right)\Gamma\left(\epsilon\right)}{\Gamma\left(2-\epsilon\right)}(-z)^{\epsilon-1}\left(1-z\right)^{1-\epsilon}+\Gamma(\epsilon-1)\right\}\nonumber\\
                                   & = &\pi^{2-\epsilon}\frac{(-m^2)^{1-\epsilon}}{(p\cdot n)}\frac{\Gamma(\epsilon)}{(1-\epsilon)}\left \{\left(1-\frac{1}{z}\right)^{1-\epsilon}-1\right\},
\eea
where in the above we have also used $D=4-2\epsilon$, to work out the limit $\epsilon \to 0$.

Let us write it as
\bea
T^\prime_D(-1,-1,0;z) & = & -\frac{m^2}{p\cdot n}\pi^2 \Omega(\epsilon),
\eea
where, with $z^{-1} \equiv w = \frac{2 p\cdot n p\cdot n^*}{m^2 n\cdot n^*}$
\bea
\Omega(\epsilon) & \equiv & \lim_{\epsilon \to 0} \frac{\Gamma\left(\epsilon\right)}{\left(1-\epsilon\right)}\left\{(1-w)\left(\pi m^2 w-\pi m^2\right)^{-\epsilon} -\left(-\pi m^2\right)^{-\epsilon}\right\} \nonumber\\
                                & = & \lim_{\epsilon \to 0} \Gamma\left(\epsilon\right)\left\{-w - \epsilon A(w) + {\cal{O}}(\epsilon{^2})\right\}, 
\eea
with
\bea
A(w) &\equiv& w-w \ln\left(-\pi m^2\right)+\left(1-w\right) \ln\left(1-w\right).
\eea

Then, 
\bea
T^\prime_D(-1,-1,0;z) & = & \int  \frac{d^{D}\!k}{\left[ (k-p)^2-m^2\right]}\frac{1}{(k\cdot n)}\nonumber \\
                                   & = & \left(\frac{2p\cdot n^*}{n\cdot n^*}\right)\pi^2 \left\{\frac{1}{\epsilon} +F(p,\,m^2)+{\cal O}(\epsilon)\right\}
\eea

Therefore, we have exactly the same divergent piece $I_{\rm div.}$ as that calculated with ML prescription and the finite piece is 
\bea
F(p,\,m^2) & = & 1\!-\!\gamma\! -\! \ln \left(-\pi m^2\right) + \frac{m^2 n\!\cdot\! n^*}{2p\!\cdot\! n p\!\cdot\! n^*}\left(1-\frac{2p\!\cdot\!n p\!\cdot\!n^*}{m^2n\!\cdot\!n^*}\right)\ln \left(1-\frac{2p\!\cdot\!n p\!\cdot\!n^*}{m^2n\!\cdot\!n^*}\right).
\eea

Finally, let us check the other solution, equation (\ref{scalarw}), in the same limit.
\bea
T^\prime_D(-1,-1,0;w) & = &  \pi^{D/2}\frac{(-m^2)^{D/2-1}}{(p\!\cdot \!n)}\!\left \{ -\Gamma\left(2\!-\!D/2\right)w\,{}_2F_{1}\left(1,2\!-\!\left.D/2;2\right|w\right) \right\}.
\eea

Introducing as before, $D=4-2\epsilon$ we rewrite it as
\bea
T^\prime_D(-1,-1,0;w) & = &  \pi^{2-\epsilon}\frac{(-m^2)^{1-\epsilon}}{(p\!\cdot \!n)}\!\left \{ -w\Gamma\left(\epsilon\right)\,{}_2F_{1}\left(1,\,\epsilon;\,2\,|\,w\right) \right\}.
\eea

We need to work out the expansion for the hypergeometric function ${}_2F_{1}\left(1,\,\epsilon;\,2\,|\,w\right) $ for $\epsilon \to 0$. To do this, first we use the identity \cite{GR}
\bea
\label{HGID}
{}_2F_1(\alpha, \beta; \gamma\,|w) & = & {\mathbb A}\,{}_2F_1(\alpha, \beta; \alpha+\beta-\gamma+1\,|1-w)\nonumber\\
&+& {\mathbb B}\,{}_2F_1(\gamma-\alpha,\gamma-\beta; \gamma-\alpha-\beta+1\,|1-w),
\eea
with
\bea
\mathbb A & \equiv & \frac{\Gamma(\gamma)\Gamma(\gamma-\alpha-\beta)}{\Gamma(\gamma-\alpha)\Gamma(\gamma-\beta)} \\
\mathbb B & \equiv & (1-w)^{\gamma-\alpha-\beta}\frac{\Gamma(\gamma)\Gamma(\alpha+\beta-\gamma)}{\Gamma(\alpha)\Gamma(\beta)}.
\eea

Then working out the expansion for the hypergeometric function  ${}_2F_{1}\left(1,\,\epsilon;\,2\,|\,w\right)$ we get
\bea
{}_2F_1(1, \epsilon; 2\,|w) & = & \frac{\Gamma(1-\epsilon)}{\Gamma(2-\epsilon)}{}_2F_1(1, \epsilon; \epsilon\,|1-w)\nonumber\\
&+& (1-w)^{1-\epsilon}\frac{\Gamma(\epsilon-1)}{\Gamma(\epsilon)}\,{}_2F_1(1,2-\epsilon; 2-\epsilon\,|1-w).
\eea

Since ${}_2F_1(a, b; b\,|1-w)=w^{-a}$ and $\Gamma(\epsilon)=(\epsilon-1)\Gamma(\epsilon-1)$ we get
\bea
\label{HG1}
{}_2F_{1}\left(1,\,\epsilon;\,2\,|\,w\right) &=&1+\epsilon \left\{ 1+\frac{(1-w)}{w}\ln (1-w)\right\}+{\cal O}(\epsilon^2),
\eea
and we arrive again at
\bea
\label{newresult}
T^\prime_D(-1,-1,0;w) & = & \left(\frac{2p\cdot n^*}{n\cdot n^*}\right)\pi^2 \left\{\frac{1}{\epsilon} +F(p,\,m^2)+{\cal O}(\epsilon)\right\},
\eea
in complete agreement with the previous result.

\subsection{Vector case}

For this case, since the expression for the $z$ solution is lengthier to deal with, we work out in detail only the shorter $w$ solution, equation (\ref{answerw}):
\bea
T^{\prime \mu}_D(w) & = & \pi^{D/2}\left\{ p^\mu{\mathbb C}^b(p)\:{}_2F_1(-j, -i-j-D/2; 1-j+l\,|w)\right. \nonumber \\
                                  & &\hspace{.75cm} -n^\mu {\mathbb C}^b(n)\:{}_2F_1(-j+1, -i-j-D/2; 2-j+l\,|w) \nonumber\\
                                  & &\hspace{.75cm} \left. -n^{* \mu} {\mathbb C}^b(n^*)\:{}_2F_1(-j, -i-j-D/2-1;-j+l\,|w) \right\}.
\eea

For the special case $i=j=-1$ and $l=0$, we have
\bea
T^{\prime \mu}_D(w) & = & \pi^{D/2}\left\{ p^\mu{\mathbb C}^b(p|-1,-1;0)\:{}_2F_1(1, 2-D/2; 2\,|w)\right. \nonumber \\
                                  & &\hspace{.75cm} -n^\mu {\mathbb C}^b(n|-1,-1;0)\:{}_2F_1(2, 2-D/2; 3\,|w) \nonumber\\
                                  & &\hspace{.75cm} \left. -n^{* \mu} {\mathbb C}^b(n^*|-1,-1;0)\:{}_2F_1(1, 1-D/2;1\,|w) \right\}, 
\eea
with 
\bea
{\mathbb C}^b(p|-1,-1;0) & = & \frac{p\cdot n^*}{n\cdot n^*}(-m^2)^{D/2-2}\Gamma(2-D/2) \nonumber\\
{\mathbb C}^b(n|-1,-1;0) & = &  \frac{(p\cdot n^*)^2}{2(n\cdot n^*)^2}(-m^2)^{D/2-2}\Gamma(2-D/2) \\
{\mathbb C}^b(n^*|-1,-1;0) & = & \frac{m^2}{2n\cdot n^*}(-m^2)^{D/2-2}\Gamma(1-D/2)\nonumber.
\eea

Introducing $D=4-2\epsilon$ as before, we have
\bea
T^{\prime \mu}_D(w) & = & \pi^{2-\epsilon}\left\{ p^\mu{\mathbb C}^b(p|-1,-1;0)\:{}_2F_1(1, \epsilon; 2\,|w)\right. \nonumber \\
                                  & &\hspace{.75cm} -n^\mu {\mathbb C}^b(n|-1,-1;0)\:{}_2F_1(2, \epsilon; 3\,|w) \nonumber\\
                                  & &\hspace{.75cm} \left. -n^{* \mu} {\mathbb C}^b(n^*|-1,-1;0)\:{}_2F_1(1, \epsilon-1;1\,|w) \right\},
\eea
with 
\bea
{\mathbb C}^b(p|-1,-1;0) & = & \frac{p\cdot n^*}{n\cdot n^*}(-m^2)^{-\epsilon}\Gamma(\epsilon) \nonumber\\
{\mathbb C}^b(n|-1,-1;0) & = &  \frac{(p\cdot n^*)^2}{2(n\cdot n^*)^2}(-m^2)^{-\epsilon}\Gamma(\epsilon) \\
{\mathbb C}^b(n^*|-1,-1;0) & = & \frac{m^2}{2n\cdot n^*}(-m^2)^{-\epsilon}\Gamma(\epsilon-1)\nonumber.
\eea

Our result can be expressed as 
\bea
T^{\prime \mu}_D(w) & = & \pi^2\left\{p^\mu \frac{p\cdot n^*}{n\cdot n^*}{\mathbb P}(\epsilon)-n^\mu \frac{(p\cdot n^*)^2}{2(n\cdot n^*)^2}{\mathbb N(\epsilon)} - n^{*\mu}\frac{m^2}{2n\cdot n^*}{\mathbb N}^*(\epsilon)\right\},
\eea
with
\bea
{\mathbb P}(\epsilon) & = & (-\pi m^2)^{-\epsilon}\Gamma(\epsilon)\,{}_2F_1(1, \epsilon; 2\,|w) \nonumber\\
{\mathbb N}(\epsilon) & = &  (-\pi m^2)^{-\epsilon}\Gamma(\epsilon)\,{}_2F_1(2, \epsilon; 3\,|w)  \\
{\mathbb N^*}(\epsilon) & = & (-\pi m^2)^{-\epsilon}\frac{\Gamma(\epsilon)}{(\epsilon-1)}\,{}_2F_1(1, \epsilon-1; 1\,|w)\nonumber.
\eea

The first hypergeometric function has already been worked out in the limit $\epsilon \to 0$ in (\ref{HG1})and the third one is a power series ${}_2F_1(1, \epsilon-1; 1\,|w) = (1-w)^{1-\epsilon}$. We need to work out the second hypergeometric function in the desired limit. 

We do it in a similar way as we have done before, using the identity (\ref{HGID}):
\bea
\label{HG2}
{}_2F_1(2, \epsilon; 3\,|w) & = & \frac{2\Gamma(1-\epsilon)}{\Gamma(3-\epsilon)}\,{}_2F_1(2, \epsilon; \epsilon\,|1-w) \nonumber\\
                                          & + & (1-w)^{1-\epsilon}\frac{2\Gamma(\epsilon-1)}{\Gamma(\epsilon)}\,{}_2F_1(1, 3-\epsilon; 2-\epsilon\,|1-w) \nonumber\\
                                          & = & \frac{2\Gamma(1-\epsilon)}{\Gamma(3-\epsilon)}w^{-2}\nonumber\\
                                          & + & (1-w)^{1-\epsilon}\frac{2}{(\epsilon-1)}{}_2F_1(1,3-\epsilon; 2-\epsilon\,|1-w)
\eea

In the above expression, we employ the identity for the hypergeometric function  \cite {GR} ${}_2F_1(a, b; c\,|z) = (1-z)^{c-a-b}{}_2F_1(c-a, c-b; c\,|z)$ that yields a terminating series at the second term as
\bea
{}_2F_1(1,3-\epsilon; 2-\epsilon\,|1-w) & = & w^{-2}\,{}_2F_1(1-\epsilon,-1; 2-\epsilon\,|1-w) \nonumber\\
                                                             & = &  \frac{1}{w^2}\left\{1-\frac{(1-\epsilon)}{(2-\epsilon)}(1-w)\right\}.
\eea

Plugging this result into (\ref{HG2}) we get, after doing expansions in power series in $\epsilon$,
\bea
{}_2F_1(2, \epsilon; 3\,|w) & = & 1+\epsilon\left\{\frac{1}{2}+\frac{1}{w}-\ln(1-w)+\frac{1}{w^2}\ln(1-w) + {\cal O}(\epsilon^2)\right\}.
\eea

Thus, finally, 
\bea
T^{\prime \mu}_D(w) & = & \int \frac{d^Dk\:\:\: k^\mu}{\left[(k-p)^2-m^2\right](2k\cdot n))} \nonumber \\
                                  & = & \left\{ p^\mu\,\frac{p\cdot n^*}{n\cdot n^*}-n^\mu \frac{(p\cdot n^*)^2}{2(n\cdot n^*)^2}-n^{*\mu} \frac{p\cdot n p\cdot n^*}{(n\cdot n^*)^2}+n^{*\mu}\frac{m^2}{2n\cdot n^*} \right\}\frac{\pi^2}{\epsilon}+F^\mu(w), 
\eea
where the finite part is given by
\bea
F^\mu(w) & = & \pi^2 p^\mu \frac{p\cdot n^*}{n\cdot n^*}\left\{1-\gamma - \ln(-\pi m^2)+\frac{(1-w)}{w}\ln(1-w)\right\} \nonumber\\
               & - & \pi^2 n^\mu \frac{(p\cdot n^*)^2}{2(n\cdot n^*)^2}\left\{\frac{1}{2}-\gamma +\frac{1}{w}-\ln(-\pi m^2)+\frac{(1-w^2)}{w^2}\ln(1-w)\right\} \nonumber\\
               & - & \pi^2 n^{*\mu} \left\{\frac{p\cdot n p\cdot n^*}{(n\cdot n^*)^2}-\frac{m^2}{2n\cdot n^*}\right\}\{1-\gamma-\ln(-\pi m^2)-\ln(1-w) \}.
\eea

The divergent pole part of the vector integral agrees with the result in \cite{Leibbrandt2}.  



\begin{thebibliography}{99}

\bibitem{Leibbrandt1} G. Leibbrandt, Phys. Rev. {\bf D 30} (1984) 2167;   E.T. Newman and R. Penrose, J. Math. Phys. {\bf3} (1962) 566.

\bibitem{ML} S. Mandelstam, Nucl. Phys. {\bf B 213} (1983) 149.

\bibitem{Leibbrandt1a} G. Leibbrandt, Phys. Rev. {\bf D 29} (1984) 1699.

 \bibitem{Halliday-Ricotta} I. G. Halliday and R. M. Ricotta, Phys. Lett. {\bf B 193} (1987) 241.
 
 \bibitem{LCML} D.M.Capper, J.J.Dulwich and M.J.Litvak, Nucl. Phys. {\bf B241} (1984) 463-476; D.M.Capper, D.R.T.Jones and A.T.Suzuki, Z. Phys. C {\bf 29} (1985) 585-596; D.M.Capper, D.R.T.Jones and N.Linden, Phys. Lett. B {\bf 181} (1986) 106-110; A.T.Suzuki, C.R.Ji, PoS Proceedings of Science, Light-Cone 2019, {\bf 374} 82 (2020) DOI: https://doi.org/10.22323/1.374.0082.
                             
 \bibitem{Basseto} A. Bassetto, G. Nardelli and R. Soldati, {\em Yang-Mills Theories in Algebraic Non-Covariant Gauges}, World Scientific Publishing Co. Pte. Ltd., Singapore, (1991).
 
 \bibitem{Leibbrandt2} G. Leibbrandt, {\em Noncovariant Gauges - Quantization of Yang-Mills and Chern-Simons Theory in Axial-type Gauges}, World Scientific Publishing Co. Pte. Ltd., Singapore, (1994).
  
 \bibitem{Suzuki} A.T. Suzuki, A. G. M. Schmidt and R. Bentin, Nucl. Phys. {\bf B 537} (1999) 549; A. T. Suzuki and A.G.M. Schmidt, Progress of Theoretical Physics {\bf 103(5)} (2000) 1011; Eur. Phys. J. {\bf C 12} (2000) 361; 
                             A. T. Suzuki and A. G. M. Schmidt, Physics Letters {\bf B 494, (2000)},  332.
                             
 \bibitem{Capper} D. M. Capper, J. J. Dulwich, and M. J. Litvak, Nucl. Phys. {\bf B 241(2)} (1984) 463.

\bibitem{Dirac} P.A.M.Dirac, Rev.Mod.Phys. {\bf 21} (1949) 392.



\bibitem{GR} {\it Table of Integrals, Series and Products}, I.S.Gradshteyn and I.M.Ryzhik, 8th Edition, Edited by Daniel Zwillinger,  Academic Press, Oxford, UK (2015).












\end{thebibliography}
\end{document}